\documentclass[aps,prb,preprint,showpacs,unsortedaddress,amsmath,amssymb]{revtex4-1}

\usepackage{graphicx}
\usepackage{dcolumn}
\usepackage{bm}
\usepackage{url}
\usepackage[colorlinks=true,linkcolor=blue,citecolor=blue,urlcolor=blue]{hyperref}

\begin{document}

\title{Lattice mismatch induced ripples and wrinkles in planar graphene/boron nitride superlattices }

\author{Dinkar Nandwana}
\affiliation
{Department of Mechanical Science and Engineering, University of Illinois at Urbana--Champaign, Urbana IL 61801, USA}
\author{Elif Ertekin}
\affiliation
{Department of Mechanical Science and Engineering, University of Illinois at Urbana--Champaign, Urbana IL 61801, USA}
\affiliation
{International Institute for Carbon Neutral Energy Research (WPI-I$^2$CNER), Kyushu University, 744 Moto-oka, Nishi-ku, Fukuoka 819-0395, Japan}
\email[e-mail:]{ertekin@illinois.edu}

\date{\today}

\begin{abstract}
A continuum theory to describe periodic ripple formation in planar graphene/boron nitride superlattices is formulated.  Due to the lattice mismatch between the two materials, it is shown that flat superlattices are unstable with respect to ripple formation of appropriate wavelengths.  A competition between bending energy and transverse stretching energy gives rise to an optimal ripple wavelength that depends on the superlattice pitch.  The optimal wavelengths predicted by the continuum theory are in good agreement with atomic scale total energy calculations previously reported in Nandwana and Ertekin, {\it Nano Lett.} {\bf 15} 1468 (2015).  
\end{abstract}


\maketitle

\section{Introduction}  
 
The rippling, crumpling, and wrinkling of elastic materials are observed at many length scales, for example: crumpling of sheets of paper~\cite{Witten95,Witten2007}, wrinkling of skin~\cite{deGennes2003}, and the buckling of the earth's crust due to viscous stresses from the underlying mantle~\cite{TurcotteSchubert2001}. The combination of geometry and physical models that capture the wavelength, amplitude, and curvature of large deformations often gives rise to surprising behavior, such as the famous problem of the wrinkling of an elastic sheet stretched in tension but clamped from contracting laterally at its ends~\cite{CerdaMahadevan}.  When it comes to the deformation of naturally thin sheets, predicting the wavelength and amplitudes of key features has been a long-standing challenge~\cite{Pellegrino2006,Fried2011}.  The large deformations are governed by the F{\"o}ppl-von K{\'a}rm{\'a}n\cite{landaulifshitz} equations.  These non-linear equations reflect a subtle balance between stretching and bending energies in deformed membranes.  With few exceptions, they are impossible to solve analytically, necessitating instead a variational approach and/or scaling analysis.  
 
Meanwhile, the recent development of two-dimensional and quasi- two-dimensional materials such as graphene~\cite{GeimNovoselov}, boron nitride~\cite{Osterwalder2004}, and transition metal dichalcogenides~\cite{Heinz2010} now provides an opportunity to observe the effects of elastic deformation at limiting length scales even down to a single atomic layer thick.  For example, ripples are intrinsic to graphene sheets~\cite{Roth2007,Geim2008}, and are expected to strongly influence the electronic~\cite{Vozmediano2008,Miranda2008}, mechanical~\cite{Galiotis2009,Fasolino2009}, and thermal~\cite{Hakonen2014} properties.  Control over the wavelength, amplitude, and orientation of one and two dimensional periodic ripples has been demonstrated~\cite{Lau2009}, and it has been suggested that rippling can give rise to a negative thermal expansion coefficient~\cite{Rybicki2014}.  Also, thermal strain engineering and scanning tunneling microscopy of suspended nanographene membranes have suggested the breakdown of continuum theories when ripple wavelengths are close to the lattice constant~\cite{Biro2012}.

\begin{figure}
  \includegraphics[width=15.24cm]{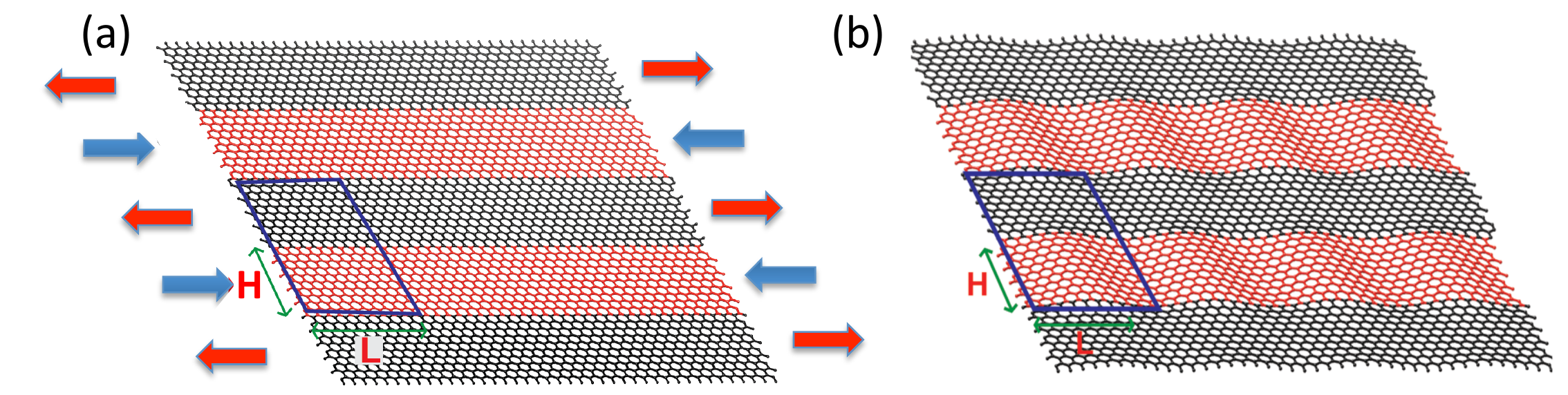}
  \caption{(a) A flat planar superlattice of composed of regular, alternating domains of graphene (black) and boron nitride (red). Due to the mismatch in the lattice constants (2.46 {\AA} and 2.52 {\AA}, respectively), when atomic registry is maintained across the interfaces the flat superlattice is composed of alternating regions of tension and compression in the direction parallel to the interface. (b) Much of the lattice mismatch strain can be accomodated by the formation of a ripple network, in which the ripples of larger amplitude in boron nitride decay to ripples of smaller amplitude in graphene.}
 \label{sch:intro}
\end{figure}

Our recent work focused on the prediction of the energetically favorable configurations of graphene/boron nitride planar superlattices~\cite{NandwanaErtekin2015}.  In these superlattices, a single atomic plane is comprised of perfectly ordered, alternating regions of boron nitride and graphene.  As shown in Fig.~\ref{sch:intro}a, due to the lattice mismatch between the two materials, the superlattice when flat is comprised of alternating regions of tension and compression along the direction parallel to the interface.  While strain-relieving misfit dislocations are expected to relieve the lattice mismatch in superlattices that are constrained to remain flat, out-of-plane buckling and ripple formation was found to also be a very effective strain relief avenue~\cite{NandwanaErtekin2015}. 

In this work, we present a detailed continuum formulation and scaling analysis of ripple formation in these superlattices.  Depending on the superlattice pitch $H$, the lattice mismatch renders the planar flat superlattices unstable with respect to the formation of a ripple network as shown in Fig.~\ref{sch:intro}b.  The ripples exhibit larger amplitude in boron nitride to accommodate its larger length, but decay to smaller amplitude in graphene.  We find that an optimal set of ripple wavelengths emerges from a competition between the ripple bending energy (favoring short wavelengths) and the transverse stretching energy due to ripple decay (favoring long wavelengths).  Our continuum and scaling analysis is in excellent agreement with atomistic models of ripple formation~\cite{NandwanaErtekin2015}, demonstrating that when properly formulated, continuum theories can hold quite nicely even at these length scales.  Although the analysis presented here is specific to graphene/boron nitride superlattices, the framework is general and can be applied to other planar superlattices.  
 
\section{Preliminary Discussion of Rippling Distortions}

\subsection{Total elastic energy of a deformed membrane.} 

In general, the total elastic energy of a rippled sheet $\mathcal{U} = \mathcal{U}_S + \mathcal{U}_B$ contains both stretching and bending contributions~\cite{landaulifshitz}: 
\begin{eqnarray} 
\label{fvk}
\mathcal{U}_S & = & \int u_S dA  = \int  \frac{1}{2}  C_{ijkl} \epsilon_{ij} \epsilon_{kl} dA  \hspace{2em}, \\
\mathcal{U}_B & = & \int u_B dA  = \int  \frac{1}{2} \kappa (\nabla^2 w)^2 dA \hspace{2em}.
\end{eqnarray}
In the stretching term $\mathcal{U}_S$, $C_{ijkl}$ are elastic constants and $\epsilon_{ij} = \epsilon_{ij}^d + F_{ij}$ are total strains, where $\epsilon_{ij}^d = \frac{1}{2}( \partial_i u_j+ \partial_j u_i)$ is the distortion strain and $F_{ij} = \frac{1}{2} (\partial_i u_k  \partial_j u_k + \partial_i w  \partial_j w)$ is the bending strain.  The indices $i,j,k$ vary over $x,y$; $dA = dxdy$; and repeated indices are summed.  The displacement field consists of in-plane ${\bf u}(x,y) = (u_x(x,y),u_y(x,y))$ and out-of-plane $w(x,y)$ contributions.  With large deflections, the nonlinear term $F_{ij}$ is non-negligible.   The bending term $\mathcal{U}_B$ reflects contributions to the energy associated with changes to the gaussian curvature $H \approx \nabla^2 w$ of the sheet, where $\kappa$ is the bending rigidity. 

It is important to note some distinctions in terminology that will be used.  The stretching (bending) energy given by $\mathcal{U}_S$ ($\mathcal{U}_B$) denotes the total energy energy contained within an area $A$ in the superlattice.  The stretching (bending) energy density is instead denoted by {\it lowercase} letters $u_S$ ($u_B$), and can vary with position.  The average stretching (bending) energy density in the superlattice is $\mathcal{U}_S/A$ ($\mathcal{U}_B/A$).

\subsection{Elastic Parameters and Superlattice Geometry} 

For the numerical results presented here, we use elastic parameters obtained from atomistic force field total energy calculations: $\kappa = 1.38$ eV, and isotropic elastic constants of $C_{xxxx} = 26$ eV/\AA$^2$, $C_{xxyy} = 7.5$ eV/\AA$^2$, and $C_{xyxy} = 9.25$ eV/\AA$^2$.  These were obtained using the LAMMPS molecular simulation package~\cite{Plimpton95} and Tersoff potentials~\cite{Tersoff} with appropriate intermixing parameters to describe interfacial interactions.  Note that with this choice of potentials the $T = 0$~K lattice constants are $a_C = 2.46$~{\AA} and $a_{BN} = 2.52$~{\AA} for graphene and boron nitride respectively, giving a lattice mismatch of $f = (a_{BN} - a_C)/a_C \approx 2.4$\%, in reasonable agreement with the experimental lattice mismatch of ~2.0\%~\cite{liuin2013}.  We use these values for the lattice parameters and the elastic moduli (rather than experimental ones) in order to directly compare the continuum results presented here to atomistic simulations.

\begin{figure}
  \includegraphics[width=5.08cm]{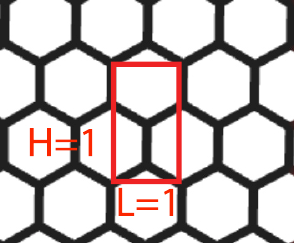}
  \caption{A single $[L,H]=[1,1]$ unit cell is outlined in red on the honeycomb lattice.  The dimensions are 2.46 {\AA}, 2.46$\sqrt{3}$ {\AA} for $L,H$ respectively.}
 \label{sch:unit}
\end{figure}

In the following analysis, all superlattices are infinitely extended in the plane.  To consider different wavelength ripples, we denote by $[L,H]$ the number of `unit cell' building blocks  which are tiled in space to build a graphene/boron nitride supercell into which a single wavelength ripple is introduced (along the $L$ direction, parallel to the interface as in Fig.~\ref{sch:intro}). Periodic boundary conditions are then imposed in all directions. A single unit cell $[L,H] = [1,1]$ is outlined in Fig.~\ref{sch:unit} which has dimensions $ 2.46 \times 2.46 \sqrt{3}$ {\AA}$^2$. In our notation, an $[L,H]$ superlattice is constructed by repeating this unit cell $L$ times parallel to the interface and $H$ times perpendicular to the interface for each subdomain (graphene and boron nitride).  Thus the area of an $[L,H]$ supercell is given by $A = 2LH(2.46^2\sqrt{3})$ {\AA}$^2$. 

\subsection{Ripple formation in an isolated sheet.}  

For simplicity we first consider a thin sheet (say, boron nitride) that has natural (unstrained) length $L(1+f)$ along the $x$-direction and natural height $H$ along the $y$-direction.  An eigenstrain $\epsilon_{ij}^o$ is introduced, in which the sheet is uniformly compressed in-plane in the $x$-direction: $u_x^o = -f x$, $\epsilon_{xx}^o = -f$.  The length of the sheet in the $x$-direction is now $L$ rather than $L(1+f)$.  The total strain energy in this reference configuration is $\mathcal{U}^{ref} = \mathcal{U}_S^{ref} + \mathcal{U}_B^{ref}$ with $\mathcal{U}_S^{ref} \approx (C_{xxxx} f^2/2) HL$, $\mathcal{U}_B^{ref} = 0$.  (Hereafter, for simplicity we will refer to $C_{xxxx}$ by $C$ so that $\mathcal{U}_S^{ref} \approx (C f^2/2) HL$).

Ripples of wavelength $\lambda$ will form if the rippled energy $\mathcal{U}$ is smaller than $\mathcal{U}^{ref}$.  Consider (i) the introduction of a ripple $w(x) = A \lambda \cos (2 \pi x/\lambda)$ of wavelength $\lambda = L$ and amplitude $A \lambda$ ($A$ is a dimensionless constant to be determined), and (ii) the superposition to the in-plane displacement field $u_x^o$ of a modulation $u_x^{mod} =\left(  \frac{f \lambda}{4 \pi} \right) \sin{\left(\frac{4 \pi x}{\lambda}\right)}$.  The modulation $u_x^{mod}$ helps relieve bending strain concentrated at the peaks and troughs of the ripples as we will show.  Now the total displacement field is  
\begin{eqnarray}
u_x & = & -f x + \left(  \frac{f \lambda}{4 \pi} \right) \sin{\left(\frac{4 \pi x}{\lambda}\right)} \hspace{2em}, \nonumber \\
w & = & A \lambda \cos{\left( \frac{2 \pi x}{\lambda} \right)} \hspace{2em}. 
\end{eqnarray}
The total strain component 
\begin{eqnarray}
\epsilon_{xx} & = & \epsilon_{xx}^d + F_{xx} \nonumber \\
& = & \frac{1}{2}( \partial_x u_x+ \partial_x u_x) + \frac{1}{2} (\partial_x w  \partial_x w + \partial_x u_x  \partial_x u_x) \nonumber \\
& = & 2 \left[ -f + A^2 \pi^2+f^2 \sin^2{(\frac{2 \pi x}{\lambda})} \right]  \sin^2{(\frac{2 \pi x}{\lambda})}
\end{eqnarray} is largely relieved everywhere when $A = \sqrt{f}/\pi$ (neglecting the smaller $f^2$ contribution).  This corresponds to the scenario that the total arc length of the rippled system recovers the natural, unstrained length of the material:
\begin{equation}
\int_0^\lambda \sqrt{1+\left( \frac{dw}{dx} \right)^2} ~dx = \frac{2 \lambda \mathcal{E}(-4A^2\pi^2)}{\pi} = \lambda (1+f) \hspace{2em},  
\end{equation}
where $\mathcal{E}$ denotes the elliptic integral of the first kind.   One can see from the series expansion $\mathcal{E} (-4 A^2 \pi^2) = \frac{\pi}{2} \left(1+\pi A^2 \right) + \mathcal{O}(A^4)$ that the choice $A = \sqrt{f}/\pi$ satisfies Eq. (5).

While the choice $A = \sqrt{f}/\pi$ substantially reduces the stretching energy from the reference configuration to a small value $\mathcal{U}_S \sim \mathcal{O}(f^4)$, ripple formation comes at the cost of a bending energy 
\begin{eqnarray}
\mathcal{U}_B & = & \frac{1}{2} \int_0^H \int_0^\lambda \kappa (\nabla^2 w)^2 dx dy \nonumber \\
& = & \frac{4 A^2 H \pi^4 \kappa}{\lambda} \nonumber \\
& = & (4 f \pi^2 \kappa) \frac{H}{\lambda}  \hspace{2em}. 
\label{bend}
\end{eqnarray} 
Thus, whether ripples of a given wavelength $\lambda$ will form depends on a competition between the energy 
$\mathcal{U}^{ref} = (C f^2/2) H \lambda$ that is offset by relieving the compression and the energy $\mathcal{U} \approx \mathcal{U}_B = (4 f \pi^2 \kappa) \frac{H}{\lambda}$ introduced by the ripple.  

With these expressions, it is possible to roughly estimate a critical wavelength for ripple formation in boron nitride due to the lattice mismatch with graphene.  By comparing the total strain energy densities of the flat {\it vs.} rippled systems ($C f^2/2$ {\it vs.} $4f\pi^2\kappa/\lambda^2$), the critical wavelength is determined to be $\lambda \approx 13.2$ \AA.  This corresponds to $\lambda \approx 5.4$ of our unit cells, and is found to be in excellent agreement with our atomistic results.  In the atomistic results, during geometry relaxation we always find that ripples with $\lambda \le 5$ unit cells relax back to a flat configuration, while those with $\lambda > 5$ unit cells are preserved.  

\section{Ripple formation in graphene - boron nitride superlattices.} 

 In the superlattices, the lattice mismatch puts the graphene and boron nitride subdomains into tension and compression, respectively (see Fig.~\ref{sch:intro}a).  To determine the optimal ripple wavelengths, we consider a periodic superlattice that consists of boron nitride ($-H < y < 0$) and graphene ($0 < y < H$) subdomains; the natural lengths of the graphene and boron nitride along the interface are $L$ and $L(1+f)$ respectively ($f$ now is the lattice mismatch, 0.024).  

We adopt a reference configuration in which the boron nitride of natural length $L(1+f)$ is kept flat, but uniformly compressed along the interfacial direction, so that it conforms perfectly to the natural graphene length $L$.  This amounts to an eigenstrain $\epsilon_{xx}^o = -f$ and corresponding displacement field $u_x^o = - f x$ for $-H < y < 0$ (with all other eigenstrain components zero everywhere else).  In this reference state, $\mathcal{U}_S^{ref} = 0$ in graphene, $\mathcal{U}_S^{ref}  \approx (C f^2/2) HL$ in boron nitride, and $\mathcal{U}_B^{ref} = 0$ everywhere.  Thus, the reference total energy for the superlattice unit cell of area $(2HL)$ is $\mathcal{U}^{ref} = (C f^2/2) HL$ and the average energy density is $\mathcal{U}^{ref}/(2HL) = C f^2/4$.  

We present two models for rippling in the heterojunctions.  In the first, the `footprint' of the superlattice along the interfacial direction is fixed at $L$ when ripples form, while in the second this constraint is relaxed.  For both models, rippling will occur if the energy of the rippled configuration $\mathcal{U} < \mathcal{U}^{ref} $. Note that the second model is more realistic for freestanding superlattices free to deform in 3D (the first model is primarily used to observe trends).  Also, for comparison purposes, we note that for superlattices constrained to stay flat, ideally the total lattice mismatch $f$ would be evenly accommodated by both materials (graphene should be stretched with $\epsilon_{xx}=f/2$ and boron nitride squished with $\epsilon_{xx}=-f/2$).  In comparison to the reference configuration, this would reduce the strain energy density to $\mathcal{U}_S/(2HL) = C f^2/8$, which is the optimal distribution of mismatch for flat systems $E^{flat}_{coh}$.  

\subsection{Model I - Fixed Footprint.}

{\it Displacements}.  In the first model, we keep the `footprint' of the rippled superlattice the same as in the reference configuration (length $L$ along the interface).  To the reference state we superpose additional displacement fields, so that the total displacement field is   
\begin{eqnarray}
w(x,y) & = & \left(\frac{A_{bn} \lambda}{1+\exp{(\frac{\beta}{H} y})} \right) \cos{\left(\frac{2 \pi x}{\lambda}\right)} \hspace{2em}, \nonumber \\ 
u_x(x,y) & = & \begin{cases} 
-f x + \left( \frac{f \lambda}{4 \pi} \right) \sin{ \left( \frac{4 \pi x}{\lambda} \right) }  , & \mbox{if } -H < y < 0  \\
 0 , & \mbox{if }  0 < y < H 
\end{cases} \hspace{1em}.
\end{eqnarray} 
The expression for $w$ corresponds to the introduction of a sinusoidal ripple of wavelength $\lambda = L$ and amplitude $(A_{bn} \lambda)/(1+\exp{(\frac{\beta}{H} y}))$.  The parameter $\beta$ sets the decay length of the ripple relative to the half-pitch $H$. If $\beta$ is small the ripple amplitude is $A_{bn} \lambda$ in boron nitride ($y<0$) and decays smoothly to zero across the interface into graphene ($y>0$); if $\beta$ is sufficiently small the ripple becomes uniform throughout both subdomains (as $\beta \rightarrow 0$, the amplitude everywhere becomes $A_{bn} \lambda/2$).  Since boron nitride is in compression but graphene is initially strain free in the reference configuration,  the former case enables rippling to relieve the strain in boron nitride without perturbing graphene.  The parameters in the variational approach ($A_{bn}$, $\beta$) are obtained by numerically minimizing the total energy $\mathcal{U} = \mathcal{U}_S + \mathcal{U}_B$.  

\begin{figure}
  \includegraphics[width=10.16cm]{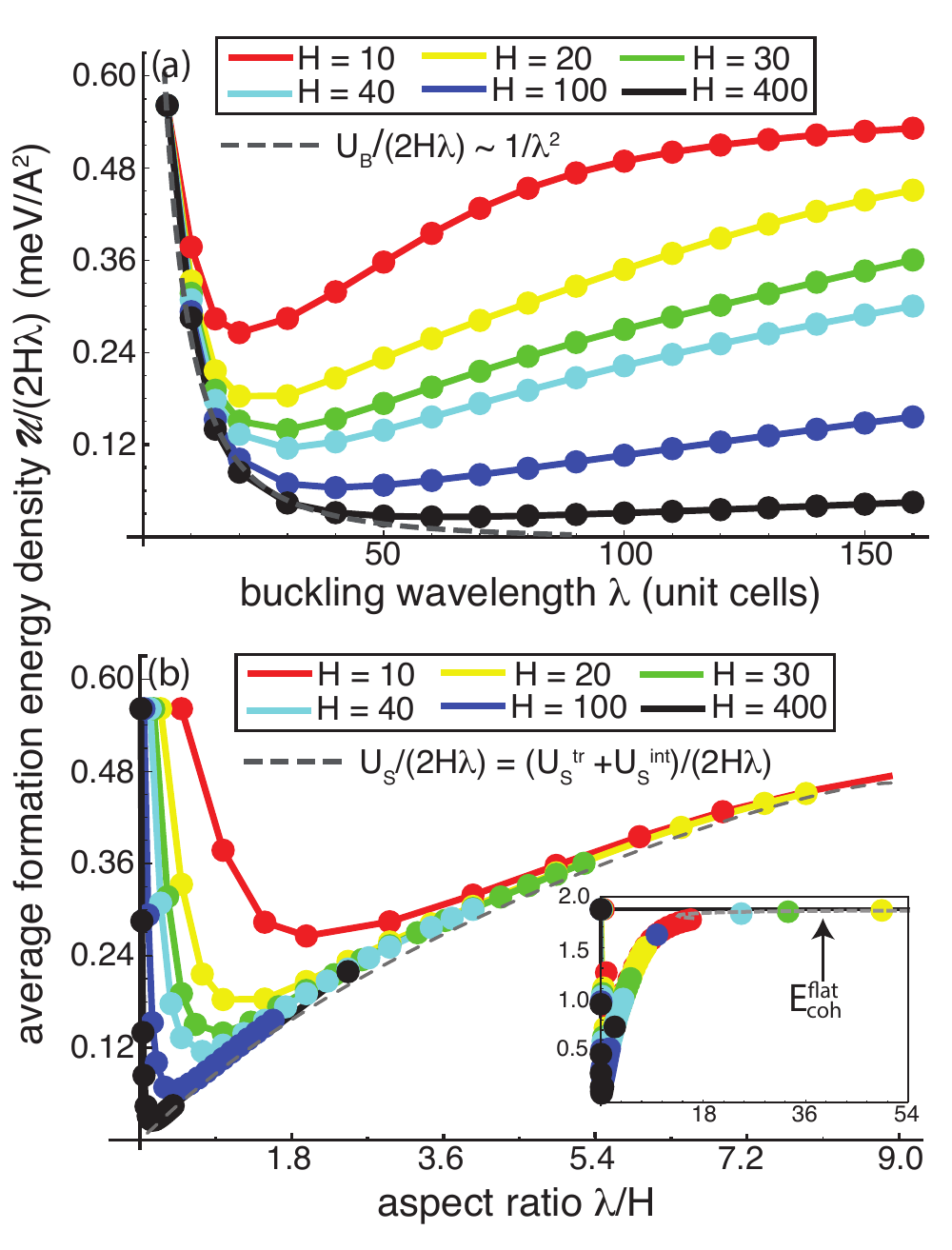}
  \caption{(a) Average formation energy density for Model I rippled superlattices as a function of ripple wavelength $\lambda$ for different half-pitch $H$.  For all $H$, initially the formation energy density drops rapidly due to the $\sim 1/\lambda^2$ nature of the bending energy.  As the wavelength increases further, the transverse stretching energy due to decay of ripples across the interface becomes dominant, inducing a transition to uniform ripples and an increase in formation energy density.  (b) Plotting the formation energy density for Model I rippled superlattices as a function of aspect ratio $\lambda/H$ collapses the curves together in the large $\lambda/H$ regime.  (Inset: as $\lambda/H$ further grows, the formation energy density increases to the value of the coherent, flat superlattices).}
 \label{sch:m1-2d}
\end{figure}

{\it Numerical Results}.  Fig.~\ref{sch:m1-2d} shows the average formation energy density $\mathcal{U}/(2H\lambda)$ computed for different rippling wavelengths $\lambda$ and superlattices of varying half-pitch $H$.  From Fig.~\ref{sch:m1-2d}a, for a given $H$, as $\lambda$ increases the average formation energy density initially drops rapidly, reaches a minimum, and then begins to increase.  The position of the minimum depends on the half pitch $H$.   For example when $H = 10$ unit cells ($\approx 43$~{\AA}), the optimal wavelength is $\lambda \approx 20$ unit cells ($\approx 50$~{\AA}).  When $H$ increases to 400 unit cells ($\approx 1704$~{\AA}), the optimal wavelength increases to $\lambda \approx 50$ unit cells ($\approx 123$~{\AA}).  Fig.~\ref{sch:m1-2d}b shows the same data, now plotted {\it vs.} superlattice aspect ratio $\lambda/H$, which appears to collapse the data onto a single curve beyond the minima (to be discussed later).  As shown in the inset of Fig.~\ref{sch:m1-2d}b, in the limit of large $\lambda/H$ the strain energy density becomes exactly the same as the optimal flat systems. In comparison to our previous atomistic results~\cite{NandwanaErtekin2015}, the trends in Fig.~\ref{sch:m1-2d} match very well to the left of the minima.  However, they differ markedly after that: in the atomistic models, no observable minimum is found, but instead the average formation energy density appears to decay quickly to a finite, converged value as $\lambda$ increases.  

{\it Contributions to Total Energy $\mathcal{U}$}.  To understand the trends in Fig.~\ref{sch:m1-2d}, we identify three key contributions to the total energy $\mathcal{U}$.  The first is the bending energy $\mathcal{U}_B$, evaluated from Eq. (2).  It is similar to Eq. (6) but now the decay of the ripple across the interface ($\partial w / \partial y$) must be included in the Laplacian $\nabla^2 w$.  As before $\mathcal{U}_B$ quickly decays as $\lambda$ increases, but (as shown later in Eq. (8)) the $\partial w / \partial y$ contribution gives rise to an additional term with scaling $1/H^2$.  Additionally, there are two dominant (and competing) stretching energy contributions: $\mathcal{U}_S \approx \mathcal{U}_S^{int} + \mathcal{U}_S^{tr}$.  The interfacial stretching energy $\mathcal{U}_S^{int}$ arises from any residual strains $\epsilon_{xx}$ in the interfacial direction once the ripples have formed.  It would be zero if a ripple in boron nitride (i) had an amplitude $A \approx \sqrt{f}/\pi$ so that the arc length recovers exactly the boron nitride natural length  $L(1+f)$, but then (ii) immediately decayed across the interface so that the graphene subdomain remains entirely strain free as well.  The second stretching energy contribution $ \mathcal{U}_S^{tr}$, is the competing term that prevents the above scenario.  It arises from strains $\epsilon_{yy}$ transverse to the interface.  The decay of the ripple across the interface gives rise to a transverse strain $\epsilon_{yy} = F_{yy} = \frac{1}{2} \left( \frac{\partial w}{\partial y} \right)^2 $.  The decay constant $\beta$ is determined by the competition between $\mathcal{U}_S^{int}$ (favoring quick decay) and $\mathcal{U}_S^{tr}$ (favoring slow decay) near the interface.  To assess the relative magnitude of these three contributions, we consider the nature of the rippling for small, intermediate, and large $\lambda/H$.  

\begin{figure}
\includegraphics[width=15.24cm]{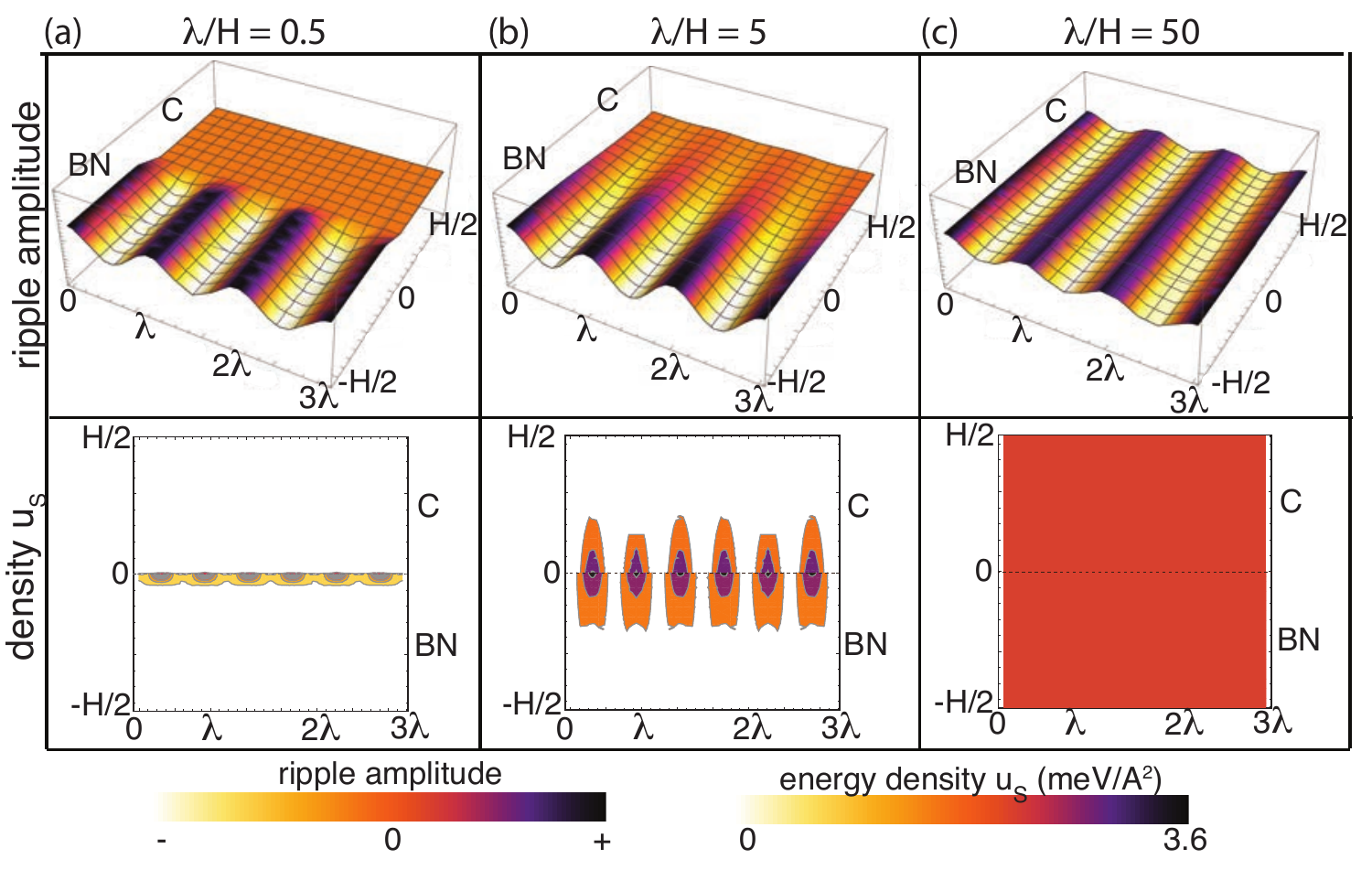}
\caption{Rippling structure and stretching energy density $u_S$ of Model I superlattices with half-pitch $H$=20 unit cells, for aspect ratios $\lambda/H$ = 0.5, 5, and 50.  For small $\lambda/H$ the ripples are present in boron nitride and decay into graphene.  At intermediate $\lambda/H$, rippling becomes somewhat preserved across the graphene subdomain. The stretching energy density $u_S$ is no longer localized to the interface.  For the largest $\lambda/H$ the ripple amplitude is almost uniform across the two materials, and is such that the mismatch strain is equally shared by them.  In this limit, the boron nitride is in compression and the graphene in tension, very similar to the flat superlattices, and the formation energy density is the same as in the flat case.  }
\label{sch:m1-3d}
\end{figure}

{\it Small $\lambda/H$ Behavior}.  In Fig.~\ref{sch:m1-2d}a, the dotted grey line plots the bending energy density $u_B$ approximated simply by $\mathcal{U}_B/(2H \lambda)$ where the expression for $\mathcal{U}_B$ is directly taken from Eq. (\ref{bend}).  The $1/\lambda^2$ decay captures the initial drop in the formation energy density very well (particularly for large $H$), suggesting that for small $\lambda/H$, most of the strain energy density comes from the bending contribution.  Our variational results also confirm this: an example of a rippled geometry and the associated stretching energy density is shown in Fig.~\ref{sch:m1-3d}a for $\lambda/H=0.5$, $H = 20$ unit cells.  The geometry shows a completely rippled boron nitride, but the ripples quickly decay to zero across the interface into graphene. The corresponding stretching energy density distribution $u_S$ is small and localized to the interfacial region. The localized nature can be understood as follows.  The ripples in boron nitride completely relieve the lattice mismatch (numerically we find $A_{bn} = \sqrt{f}/\pi$), leaving the interior strain-free.  The decay constant $\beta$ is such that the ripple decays quickly and the interior of graphene is also strain free.  By contrast, at the interface where the ripple amplitude is changing, there is a residual strain $\epsilon_{xx}$ that provides a small but non-zero $u_S^{int}$ and a transverse strain $\epsilon_{yy}$ that provides a small but non-zero $u_S^{tr}$.  On the whole however, for small $\lambda/H$ the preferred geometry is one in which ripples form only in boron nitride to relieve the mismatch strain, but then quickly decay into graphene.  The residual average strain energy density is mainly composed of a bending contribution $u_B$.

{\it Intermediate $\lambda/H$ Behavior}.  To explain the increase of the average formation energy density in Fig.~\ref{sch:m1-2d} to the right of the minima, we consider how the stretching energy densities $u_S^{tr}$ and $u_S^{int}$ vary with aspect ratio $\lambda/H$.  If the ripple nature were to remain the same as for small $\lambda/H$, the amplitude of the ripple in boron nitride ($A_{bn} \lambda = \sqrt{f} \lambda / \pi$) would increase linearly with $\lambda$, but then still must quickly decay to zero into the graphene.  This decay of large amplitude wave over a relatively short region would result in an increased transverse stretching energy.  The outcome of this competition is seen in Fig.~\ref{sch:m1-3d}b: for $\lambda/H = 5$ to avoid this energy penalty, the rippling geometry is different.  The ripples in boron nitride decay somewhat, but persist throughout the graphene subdomain.  While this smoothening of ripples avoids the large transverse stretching energy penalty, the interior of the two subdomains are no longer strain-free (the strain energy $\mathcal{U}_S^{int}$ is larger now due to residual lattice mismatch $\epsilon_{xx} = \epsilon_{xx}^d + F_{xx}$).  Thus, in the intermediate $\lambda/H$ regime ($\lambda/H \approx 5$), the parameters $(A_{bn},\beta)$ are determined as an optimal compromise between $\mathcal{U}_S^{tr}$ and $\mathcal{U}_S^{int}$.  This is also evident in Fig.~\ref{sch:m1-3d}b, where the local stretching energy density $u_S$ is found to be larger, and now extends further away from the interface.

{\it Large $\lambda/H$ Behavior}.  At large $\lambda/H = 50$ (Fig.~\ref{sch:m1-3d}c), the ripples are now almost entirely uniform throughout both subdomains.  This suggests that at large $\lambda/H$, the ripple geometry is dictated by avoiding transverse stretching energy.  In fact, in Fig.~\ref{sch:m1-3d}c numerically we find that $u_S^{tr} \approx 0$ and $u_S^{int} \approx Cf^2/8$, mimicking the distribution of the optimal flat superlattices.  Accordingly, we find numerically that amplitude $A_{bn} = \sqrt{f}/\pi$, and decay $\beta \rightarrow 0$ so that the lattice mismatch $f$ is shared evenly by a compressive strain of $\epsilon_{xx} = -f/2$ in boron nitride and a tensile strain of $\epsilon_{xx} = f/2$ in graphene.  (The fixed footprint prevents relaxation to a flat configuration, but even so the stretching energy density is the same). This is consistent with the inset of Fig.~\ref{sch:m1-2d}b, in which the formation energy density was observed to increase back to the value of the flat system.  

\subsection{Model I - Scaling Analysis}

Using our variational expressions for the displacement fields, mathematical expressions for the average bending energy density $\mathcal{U}_B/(2H\lambda)$ and the average interfacial and transverse stretching energy density $\mathcal{U}_S^{int}/(2H\lambda)$ and $\mathcal{U}_S^{tr}/(2H\lambda)$ can be obtained.  This is useful to extract the scaling nature of each contribution.  We obtain 
\begin{eqnarray}
\mathcal{U}_B/(2H\lambda) & = & \frac{1}{2H\lambda} \int_{-H}^H \int_0^\lambda \kappa (\nabla^2 w)^2 dx dy \nonumber \\
& = & \left( \frac{1}{\lambda^2} \right) (A_{bn}^2 \kappa) C_1 +  \left( \frac{1}{H^2} \right) (A_{bn}^2 \kappa)    \left[ C_2 + C_3 \left( \frac{\lambda}{H} \right)^{2}  \right]  \\ 
\mathcal{U}_S^{tr}/(2H\lambda) & = &   \left( \frac{\lambda}{H} \right)^{4} (A_{bn}^4 C)  C_4 \\
\mathcal{U}_S^{int}/(2H\lambda) & = & (A_{bn}^4 C) C_5 + C \mathcal{F}(A_{bn},f,\beta)
\end{eqnarray}
where each $C_i$ $(i=1,...5)$ is a function of $\beta$, and $\mathcal{F}$ is a non-separable function of $\beta,f,A_{bn}$.  The explicit forms of the $C_i$ and $\mathcal{F}$ are not provided here -- they are straightforward to obtain but tedious.  Most important are the scaling of the functions with $\beta$, which have leading order terms as:  $C_1 \sim \beta^0$, $C_2 \sim \beta^4$, $C_3 \sim \beta^6$, $C_4 \sim \beta^4$, $C_5 \sim \beta^0$, $\mathcal{F} \sim \beta^0$. The two terms in Eq. (10) correspond to contributions from graphene and boron nitride, respectively.  Since in boron nitride buckling is favorable, $\mathcal{F}$ is a function that approaches zero when the amplitude $A_{bn} \approx \sqrt{f}/\pi$ and the decay parameter $\beta$ is large.  Relatedly, the function $C_5$ approaches zero for sufficiently large $\beta$ (or when $A_{bn} \approx 0$) since rippling is not favorable in graphene.  

We note several observations from these expressions.  The average bending energy density has explicit $1/\lambda^2$ and $1/H^2$ dependences.  By contrast, the average transverse stretching energy density has no explicit dependence on $L$ or $H$, but only depends on the aspect ratio $(L/H)^4$.  Interestingly, the interfacial stretching term also has no explicit dependence on $\lambda$ or $H$, nor does it depend on aspect ratio $\lambda/H$.  This  explains the collapse of the numerical results to the right of the minimum in Fig.~\ref{sch:m1-2d}b (where the bending contribution has become negligible).  For such systems with negligible $\mathcal{U}_B$, the $(A_{bn},\beta)$ that minimize Eq. (9) and (10) can also be numerically determined.  We have plotted this line on Fig.~\ref{sch:m1-2d}b as well, and it corresponds well to the previous numerical results.  The scaling of the two stretching energy terms, (i) $\mathcal{U}_S^{tr}/(2H\lambda) \sim A_{bn}^4 \beta^4 (\lambda/H)^4$, and (ii) $\mathcal{U}_S^{int}/(2H\lambda)$ which comprises of two terms $\sim A_{bn}^4$ and $\sim (A_{bn}-2\sqrt{f}/\pi)^4$  for graphene and boron nitride respectively, are consistent with smoothening of ripples as $\lambda/H$ grows.  In the limit $\lambda/H \rightarrow \infty$, Equations (9) and (10) are minimized by $\beta \rightarrow 0$, $A_{bn} \rightarrow \sqrt{f}/(\pi)$ for which $\mathcal{U}_S^{tr}/(2H\lambda) \approx 0$ and $\mathcal{U}_S^{int}/(2H\lambda) \approx Cf^2/8$, recovering again the same strain energy distribution of the optimal flat systems.  

\subsection{Model II - Variable Footprint.}

The first model does not properly capture our atomistic results in which, as the wavelength $\lambda$ increases, the formation energy drops initially but then appears to reach a steady value rather than increase~\cite{NandwanaErtekin2015}.  Therefore, we introduce a different variational form for the displacement fields, designed to reduce the transverse stretching energy $\mathcal{U}_S^{tr}$ in order to maintain non-uniform rippling even as the wavelength $\lambda$ grows.  For instance, our atomistic simulations show that for $\lambda/H = 5$ as in Fig.~\ref{sch:m1-3d}b, both materials are rippled to large amplitude, but the boron nitride subdomain rippling amplitude is somewhat larger to accommodate its longer length.  

Therefore, in Model II, we allow the `footprint' of the superlattice to be squeezed from its reference value $L$ to some other value $L(1-f_s)$ in the interfacial direction, making it advantageous for both the graphene and the boron nitride to form ripples.  Consequently, as we will show, the difference in the ripple amplitude needed in both materials to relieve lattice mismatch is reduced, as is the transverse stretching energy $\mathcal{U}_S^{tr}$.  We find that this delays the onset of the increase in the formation energy density with $\lambda/H$ in comparison to Model I, and predicts formation energies in very good agreement with our atomistic results.  An important prediction however is that the formation energy density of the rippled superlattices in our atomistic results would  eventually also increase as $\lambda/H$ increases.  Unfortunately this occurs at very large wavelengths that render the supercells computationally prohibitive for atomistic simulation  -- so we are unable to directly verify this prediction.  

{\it Displacements}. In Model II, the variational form of the displacement fields are  
\begin{eqnarray}
w(x,y) & = & \left[ A_{c} + \left( \frac{A_{bn} - A_c}{1+\exp{\left(\frac{\beta}{H} y \right)}} \right) \right]  \lambda \cos{\left(\frac{2 \pi x}{ \lambda }\right)} \hspace{2em}, \\ 
u_x(x,y) & = & \begin{cases}
\left(-f -f_s \right) x + \left( \frac{(f+f_s) \lambda}{4 \pi} \right) \sin{ \left( \frac{4 \pi x}{\lambda} \right) } , & -H < y < 0 \\
- f_s x + \left( \frac{ f_s \lambda}{4 \pi} \right) \sin{ \left( \frac{4 \pi x}{\lambda} \right) } , &   0 < y < H 
\end{cases} \hspace{1em}.
\end{eqnarray} 
The ripple amplitude now smoothly decays from $A_{bn} \lambda$ in  boron nitride to $A_{c} \lambda$ in graphene.  Additionally, the footprint of the superlattice is now $\lambda = L (1-f_s)$, and now there are four variational parameters: $f_s > 0$, the amplitudes $A_c$, $A_{bn}$, and the decay constant $\beta$.  

With this set of reference displacements, graphene is subjected to an in-plane compressive strain of $\epsilon_{xx} \approx -f_s$, and boron nitride is subjected to an in-plane compressive strain of $\epsilon_{xx}   \approx -f-f_s$.  The advantage is that amplitudes $A_c$, $A_{bn}$ can be chosen to optimally relieve this in-plane compression, at lower transverse stretching energy cost.  The criterion that the arc length of each subdomain equal each material's natural length becomes: 
\begin{eqnarray}
\int_0^\lambda \sqrt{1+\left( \frac{dw}{dx} \right)^2} ~dx & = & \frac{2 \lambda \mathcal{E}(-4A_{bn}^2\pi^2)}{\pi} = L (1+f) \approx \lambda \left( 1+f + f_s \right) \hspace{1em};  \nonumber \\
\int_0^\lambda \sqrt{1+\left( \frac{dw}{dx} \right)^2} ~dx & = & \frac{2 \lambda \mathcal{E}(-4A_{c}^2\pi^2)}{\pi} = L \approx \lambda \left(  1+ f_s \right) \hspace{2em},
\end{eqnarray} 
for boron nitride and graphene respectively.  To satisfy this, the amplitudes $A_{c}^2 \pi^2 \approx f_s$ and $A_{bn}^2 \pi^2 \approx f+f_s$.  As $f_s$ grows, then $A_{bn}$ and $A_{c}$ become more similar.  The most dramatic change to the scaling relations in Eq. (9) for $\mathcal{U}_S^{tr}$ is that now $A_{bn}^4 \beta^4 \rightarrow (A_{bn}-A_{c})^4 \beta^4$.  Thus, in Model II as $\lambda/H$ grows it is still possible to maintain large $\beta$ with small transverse stretching energy.  

\begin{figure}
  \includegraphics[width=10.16cm]{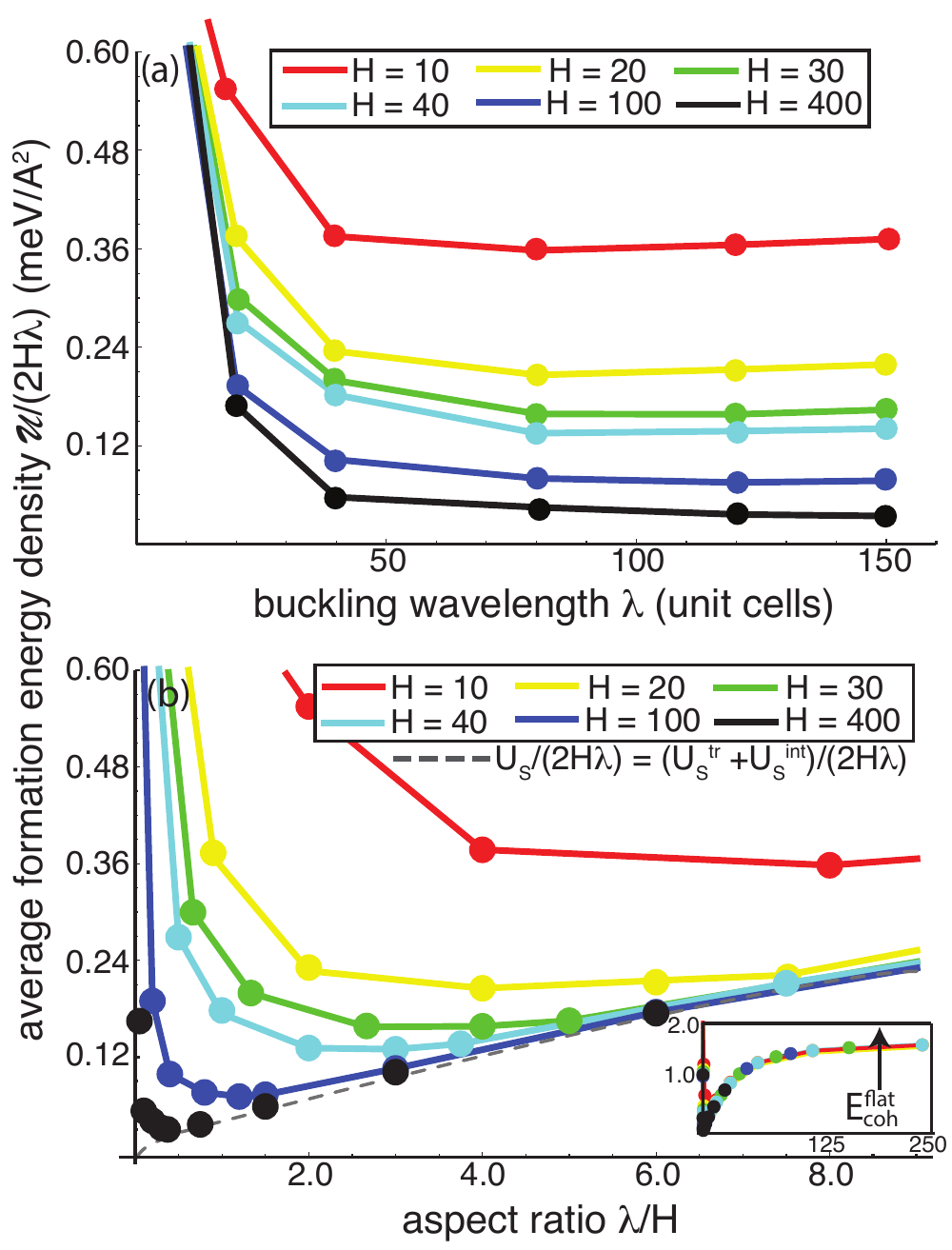}
  \caption{(a) Average formation energy density for Model II rippled superlattices as a function of ripple wavelength $\lambda = L (1-f_s)$ for different half-pitch $H$. For all $H$, (as in Model I) initially the formation energy density drops rapidly due to the $\sim 1/\lambda^2$ nature of the bending energy.  However, since both materials are now compressed along the interfacial direction, ripple formation is possible in both.  The reduced amplitude discrepancy between $A_{bn}$ and $A_c$ effectively delays the onset of the increase in the formation energy density with increasing $\lambda/H$.  (b)  The same data plotted as a function of aspect ratio $\lambda/H$; again the results collapse to the same curve for sufficiently large aspect ratio.}
 \label{sch:m2-2d}
\end{figure}

{\it Numerical Results}.  Figure~\ref{sch:m2-2d} shows the formation energy densities computed from Model II.  The behavior is now different from Model I, and both qualitatively and quantitatively matches our prior atomistic results~\cite{NandwanaErtekin2015}.  The initial sharp drop of the average formation energy density with wavelength $\lambda$ still arises from the $1/\lambda^2$ scaling of the bending energy density $\mathcal{U}_B/(2HL)$.  Now however, the formation energy density drops, remains stable for some time, and much more slowly begins to increase.  The onset of the increase is slowed by the reduced amplitude discrepancy in the two subdomains ($A_{bn} \approx \sqrt{f+f_s}/\pi$ to $A_c \approx \sqrt{f_s}/\pi$), reducing the transverse stretching energy.  Also, from Fig.~\ref{sch:m2-2d} we note that the numerically obtained data does collapse onto itself at larger $\lambda/H$ as in model I.  The optimal rippling wavelengths again depend on the half-pitch $H$: when $H = 10$ unit cells ($\approx 43$~{\AA}), the optimal wavelength is $\lambda \approx 60$ unit cells ($\approx 150$~{\AA}).  When $H$ increases to 400 unit cells ($\approx 1700$~{\AA}), the optimal wavelength increases to $\lambda \approx 160$ unit cells ($\approx 390$~{\AA}). 

\begin{figure}
  \includegraphics[width=15.24cm]{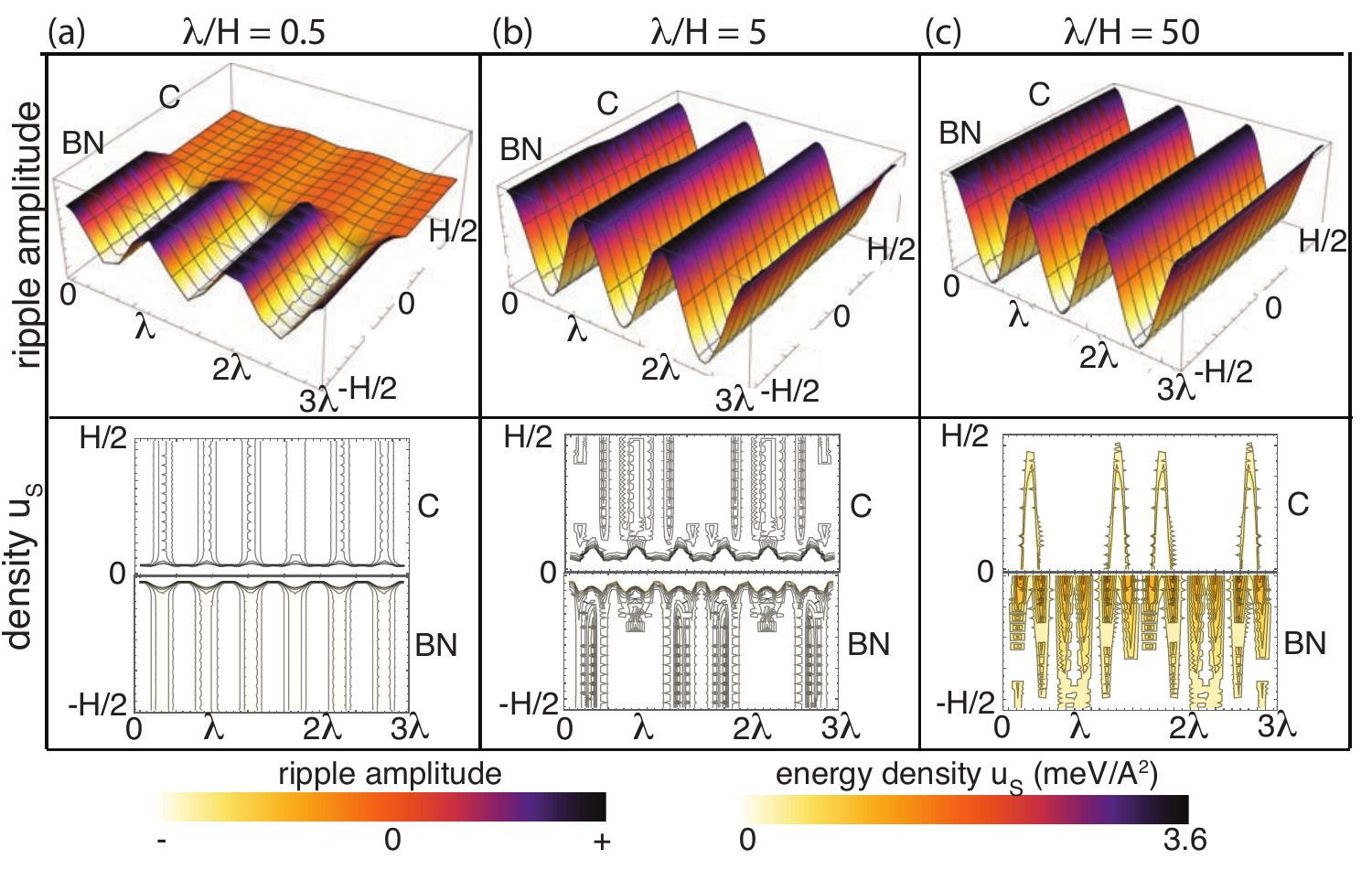}
  \caption{Rippling structure and stretching energy density $u_S$ of Model II superlattices with half-pitch $H$=20 unit cells, for aspect ratios $\lambda/H$ = 0.5, 5, and 50.  The stretching energy density $u_S$ is very small for all geometries shown (and largely composed of numerical noise -- we maintain the same scale bar as in Fig.~\ref{sch:m1-3d}, to facilitate comparison). }
  \label{sch:m2-3d}
\end{figure}

Analogously in Fig.~\ref{sch:m2-3d} the ripple geometries for $\lambda/H = 0.5, 5$ and $50$, with $H = 20$ unit cells, show first a rippled boron nitride but nearly flat graphene (as before in Model I), then a geometry in which both materials have ripples but a small drop in the ripple amplitude across the interface is observable, to finally one where the ripples {\it appear} more uniform throughout the entire domain (but in fact there remains some amplitude discrepancy, that is not perceptible because it is small relative to the total amplitude).  In all cases, the stretching energy density is small.  Although it has begun to grow at $\lambda/H = 50$, it is still far from energy density of Model I at this aspect ratio, which had already approached the optimal flat value.  

Lastly, we note that even in this model, the aspect ratio $\lambda/H$ cannot increase without bound without causing the total energy density to increase.  This limit arises from the fact that the squishing parameter $f_s$ cannot grow arbitrarily large.  Since the wave amplitudes are approximately given by $A_{bn} \approx \sqrt{f+f_s}/\pi$ and $A_c \approx \sqrt{f_s}/\pi$, then the strain component $\epsilon_{xx} = \epsilon_{xx}^d + F_{xx}$ will become large due to the nonlinear term $(\partial_x u_x)^2/2$.  This itself limits the degree of squishing ($f_s$) possible before the stretching energy becomes prohibitively large.  Numerically, we find that $f_s$ never grows beyond 0.14, at which point the lateral stretching energy starts to exceed the flat, coherent strain energy.

\section{Conclusions}
In summary, a variational continuum model and scaling analysis is developed to describe the formation of ripples in planar graphene/boron nitride superlattices.  Within periodic boundary conditions, superlattices are found to be unstable with respect to formation of ripples of appropriate wavelength which efficiently accommodate lattice mismatch strain.  A competition between the bending energy (favoring large wavelength ripples) and transverse stretching energy (favoring small wavelength ripples) gives rise to a set of optimal rippling wavelengths that minimize the total energy of the deformed superlattice.  These optimal wavelengths depend on the superlattice pitch, but vary from 15-40 nm for the geometries considered here.  In principal, the predictions presented may be validated in experiment for freestanding superlattices free to deform out of plane.  

\begin{acknowledgments}
We gratefully acknowledge discussions with Daryl C. Chrzan. We acknowledge financial support from National Science Foundation under Grant CBET-9122625. This work used the Extreme Science and Engineering Discovery Environment (XSEDE) under
Grant. No. TG-DMR120080, which is supported by National Science Foundation grant number OCI-1053575. Computational resources were also provided by the Illinois Campus Cluster program, maintained by the NCSA.
\end{acknowledgments}

\newpage 

\newpage 

\newpage 

\newpage 

\newpage 



\clearpage 
\bibliographystyle{apsrev}

\end{document}